\documentclass[12pt]{article}
\usepackage[paper=letterpaper,margin=2cm]{geometry}
\usepackage{amsmath, amssymb, amsfonts}
\usepackage{newtxtext, newtxmath}
\usepackage{enumitem}
\usepackage{tablefootnote}
\usepackage{titling}
\usepackage[colorlinks=true]{hyperref}
\usepackage{graphicx}
\usepackage{textgreek}
\usepackage{setspace} % double spacing package
\usepackage{helvet}
\usepackage{tikz}
\usepackage{caption}
\usetikzlibrary{positioning, arrows.meta, calc}
\usepackage[utf8]{inputenc}

\title{Sequential Confirmatory Factor Analysis: A Novel Approach to Latent Variable Measurement}
\date{\today\\[-1.5em]\\Word Count: 5,975}

\begin{document}

\doublespacing % <— this alone is enough

% --------------------------
% Custom Title Page (edit, not add)
% --------------------------
\begin{titlepage}
  \centering
  \vspace*{3cm}
  {\LARGE Sequential Confirmatory Factor Analysis: A Novel Approach to Latent Variable Measurement \par}
  \vspace{1.5cm}
  {\large Zachary Esses Johnson \par}
  \vspace{0.5cm}
  {\large University of Virginia \par}
  \vspace{1.5cm}
  {\large \textbf{Preprint: Not peer reviewed. Please do not cite or distribute without permission.} \par}
  \vfill
  \vspace{2cm}
  	oday
\end{titlepage}

\section*{Abstract}
Factor score estimation in small sample sizes often encounters parameter bias and convergence failures when constructing hierarchical national/sub-national indices. This paper proposes a novel method for hierarchical factor analysis called "sequential Confirmatory Factor Analysis". Instead of estimating multiple levels of factors at the same time, this approach calculates factor scores sequentially from the lowest to highest levels. This sequential estimation keeps the original sample size in each step and also removes cross-level covariance estimation. Using a series of Monte Carlo simulations, we isolate the difference between sequential Confirmatory Factor Analysis and traditional Confirmatory Factor Analysis by comparing their resulting factor scores to the true latent variables under varying conditions. We also estimate the WJP Rule of Law Index using traditional Confirmatory Factor Analysis, Bayesian Confirmatory Factor Analysis, and sequential Confirmatory Factor Analysis to test performance. Our findings demonstrate that sequential Confirmatory Factor Analysis significantly outperforms the traditional model for indices with simple/moderate complexity. Traditional Confirmatory Factor Analysis performs better where the data are skewed. Where the hierarchical model becomes complex, the two methods perform similarly. Finally, sequential Confirmatory Factor Analysis can provide valid estimates where traditional or Bayesian Confirmatory Factor Analysis fail to converge.

\newpage

\section{Introduction}
Political scientists can use Confirmatory Factor Analysis ("CFA") to test hypothesized relationships between items (i.e., instances of bribery) and latent factors (i.e., overall corruption). CFA can also be hierarchical, allowing political scientists to map relationships between items, lower order factors, and higher order factors (i.e., how constitutional safeguards are connected to freedom of speech, and by extension democracy). Unfortunately, small samples in CFA lead to parameter bias and model instability. Basic sample size thresholds range from 5-10 observations per parameter (Kline, 2011; Costello and Osborne, 2005), but more complicated Monte Carlo sample size diagnostic tools typically identify higher N dependent on model complexity (Muthen and Muthen, 2002). Hierarchical CFA requires even more observations as it estimates the relationships between items, lower order factors and higher order factors simultaneously, adding cross-level covariation estimation and overall model complexity (Sagan, 2019).

Comparativists encounter these sample size issues where they construct country-level indices (i.e., levels of democracy, corruption, freedom of speech, etc.). As country counts are inflexible, expansion is only possible through time series or subnational analyses. The only other alternatives are to omit items and/or use a different estimation tool.

Sequential CFA is our novel strategy that simplifies traditional CFA designs to make smaller-N analysis possible. Sequential CFA reduces model complexity by first fitting an initial CFA to estimate factor scores for lower-order constructs and then subsequently estimating second-level latent variables using those factor scores as observed items. This process effectively separates the hierarchical model into a two stage estimation, retaining the original sample size at each level and eliminating the need to estimate cross-level covariances that inflate standard errors in traditional CFA.

The logic of this approach draws on factor score regression, a technique where estimated factor scores from a first-stage measurement model are used as inputs for subsequent regression or structural modeling. Factor score regression is effective because the predicted factor scores preserve the variance structure of the latent variables well enough to approximate relationships that would otherwise require simultaneous estimation. By replacing latent variables with their estimated scores, researchers reduce model complexity and mitigate parameter bias in small-sample settings (Skrondal \& Laake, 2001; Croon, 2002; Cox and Kelcey, 2019).

Sequential CFA extends this logic to hierarchical measurement models, which are more methodologically demanding than typical structural regressions. In a hierarchical CFA, the goal is not to model predictive relationships between constructs but to accurately represent the latent factor structure underlying observed indicators, which introduces theoretical challenges. First, errors will propagate because first-stage factor scores include measurement error that can bias the second-stage estimation. Second, there may be identification issues because factor scores leave out joint information that traditional CFA would utilize. The above demonstrates that sequential CFA is not a re-application of factor score regression and instead attempts to resolve a greater estimation issue than that resolved by factor score regression because measurement accuracy and validity is essential to sequential CFA's utility.

Comparativists and international relations scholars will benefit from Sequential CFA. Many existing indices are hierarchical (i.e., V-Dem Index, Freedom in the World Index, WJP Rule of Law Index, etc.) and could either be re-estimated or validated using Sequential CFA where small-N restrictions would have previously eliminated this option. Furthermore, comparativists will now have an accessible hierarchical CFA estimation tool for future studies where Bayesian methods or formula-based techniques are inappropriate. As most cross-national studies are small-N for factor analysis, Sequential CFA may facilitate the construction of many more data-driven political science indices in the future. Despite the above, we acknowledge that the payoffs of Sequential CFA may have unintended consequences. Measurement error from the first CFA could propagate into the second. This study therefore investigates whether sequential CFA enables the analysis of hierarchical latent factor structures with small sample sizes (without introducing problematic measurement error) and under what conditions this result holds.

\section{Background}
\subsection{Indices in Global Politics}
Researchers commonly construct indices in global and sub-national politics to compare jurisdictions to one another. By aggregating multiple survey responses, other index scores, and economic/political indicators allows scholars to evaluate concepts that are otherwise difficult to measure directly, such as democracy, rule of law, corruption, or state capacity. Examples of such indices include the Varieties of Democracy Index (V-Dem), Freedom House's Freedom in the World Index, and the World Justice Project's Rule of Law Index.

Indices are rarely constructed with regard to measurement theory, resulting in invalid scores. Accurate and valid indices must weigh components so that they closely resemble the true unobserved latent variable. As such, researchers must carefully choose how to aggregate indicators. This requirement motivates the use of data-driven measurement techniques, such as CFA, which can directly estimate latent constructs and minimize the biases introduced by arbitrary weighting schemes.

\subsection{Where CFA Outperforms Alternative Measurement Methods}
\subsubsection{Arbitrary Weighting}
As a general rule, arbitrary approaches generally lead to significant measurement error (Bollen, 1989), and this outcome can be demonstrated formally. An index \( \bar{I} \) should reflect the underlying latent construct \( \boldsymbol{\eta} \) as much as possible. If \( \bar{I} \) is constructed from \( n \) observed variables \( x_1, x_2, \dots, x_n \) with arbitrary weights \( w_1, w_2, \dots, w_n \):

\[
\bar{I} = \frac{\sum_{i=1}^n w_i x_i}{\sum_{i=1}^n w_i}
\]

then this means that there is a correct value for each $w_i$ that is unknown. We have no insight into what $w_i$ should be through arbitrary guessing. Therefore, resorting to data-driven processes (such as CFA) to estimate latent variables is a responsible and principled option. CFA can be expressed as:

\[
\mathbf{X} = \boldsymbol{\Lambda} \boldsymbol{\eta} + \boldsymbol{\epsilon}
\]

where \( \mathbf{X} \) are items, \( \boldsymbol{\Lambda} \) is the matrix of factor loadings, \( \boldsymbol{\eta} \) are the latent variables, and \( \boldsymbol{\epsilon} \) is the vector of measurement errors.

The model's covariance matrix \( \boldsymbol{\Sigma} \) in CFA maps the expected relationships among the observed variables based on the latent structure. It is computed from the factor loading matrix \( \boldsymbol{\Lambda} \), the variance-covariance matrix of the latent variables \( \boldsymbol{\Psi} \), and the variance-covariance matrix of the measurement errors \( \boldsymbol{\Theta}_\epsilon \) (Brown, 2015; UCLA, 2025):

\[
\boldsymbol{\Sigma} = \boldsymbol{\Lambda} \boldsymbol{\Psi} \boldsymbol{\Lambda}^\top + \boldsymbol{\Theta}_\epsilon
\]

CFA compares \( \boldsymbol{\Sigma} \) with the observed covariance matrix and minimizes the difference to find the optimal model fit. Once a CFA model has been estimated, factor scores representing the latent variable can be calculated for each observation. These factor scores can then be used for further analysis (either measurement or inference). Bartholomew, Deary, and Lawn (2009, p. 576) derive Bartlett factor scores as follows:

\[
\mathbf{W} = \boldsymbol{\Gamma}^{-1} \boldsymbol{\Lambda}^\top \boldsymbol{\Theta}_\epsilon^{-1}
\]

where \( \boldsymbol{\Gamma} = \boldsymbol{\Lambda}^\top \boldsymbol{\Theta}_\epsilon^{-1} \boldsymbol{\Lambda} \), and \( \boldsymbol{\Theta}_\epsilon \) is the diagonal matrix of unique variances. The factor score is then calculated through:

\[
\hat{\boldsymbol{\eta}} = \mathbf{W} \mathbf{x}
\]

Ultimately, the above demonstrates why CFA is almost always preferable to arbitrary weighting. First, most indices are constructed because we are unsure what and how individual items contribute to the underlying latent factor. If these connections are unknown, then guessing these relationships with limited knowledge is not empirically grounded. Second, as measurement models become more complicated, researchers become less equipped to estimate the relationships between the variables themselves, leading to accumulating measurement errors. Many political science indices include dozens of individual variables, and therefore specifically benefit from a data-driven approach such as CFA.

\subsubsection{Markov Chain Monte Carlo}
Markov Chain Monte Carlo ("MCMC") may be more appropriate for hierarchical measurement modeling than CFA when dealing with complex data. Unlike CFA, which uses maximum likelihood estimation, MCMC provides full posterior distributions (which provides more information) and therefore makes it generally more accurate for small samples (Shin, Thayer, and Hwang, 2024).

However, there are drawbacks to MCMC. CFA enables researchers to test hypothesized relationships between items and underlying latent constructs in a way that is not replicable using MCMC. As well, CFA is more computationally efficient and therefore more effective when analyzing large datasets or fitting complex models (Lee and Song, 2004). CFA can also handle missing data using full information maximum likelihood estimation (a well established method), whereas MCMC requires additional estimation procedures. Finally, CFA has several fit indices to assess model suitability. Lastly, MCMC is not always able to handle models fitting data with small samples and many parameters. In contrast, MCMC relies on posterior predictive checks and other diagnostic tools that are harder to interpret for the average social scientist.

\subsubsection{Principal Components Analysis}
Principal Components Analysis (PCA) is a method that reduces dimensionality across items to create components through maximizing the explained variance. PCA is ideal for exploratory analysis, but has limited measurement application.

First, PCA transforms observables into a series of components that together explain the variance in the data. This means that it does not model latent constructs or identify any theoretical structure (Jolliffe, 2002). PCA weights are determined exclusively by statistical properties like eigenvalues of the covariance matrix. There is no flexibility to include hypotheses about the relationships within the data. Second, PCA does not distinguish between shared variance and error variance, which results in measurement error because noisy items may receive more weight than are warranted (Fabrigar et al., 1999). In turn, this leads to faulty measurements of components that do not accurately reflect the underlying data. Third, PCA assumes component orthogonality (Widaman, 1993), which is an unrealistic constraint for the most political indices. Indices in comparative politics and international relations measure concepts such as the strength of the rule of law, democracy, etc., which involve correlated dimensions. CFA instead does not demand orthogonality and allows correlated latent factors.

\section{Sequential and Traditional Factor Analysis}
CFA can model hierarchical latent structures. When measuring a concept like liberal democracy, one would need a set of observable variables such as the presence of a separation of powers, number of independent media outlets, and the existence of free and fair elections, among others. These variables as is could form a latent factor model, but given that they each theoretically contribute to different dimensions of liberal democracy, the data would be best modeled hierarchically. These items can be grouped into first-order factors (such as checks and balances, voting rights, and freedom of speech), which in turn contribute to a second-order latent construct that represents liberal democracy.

This measurement design is common in political science research where indices are composed of related subindices. However, modeling these structures using traditional CFA requires that all levels be estimated simultaneously:

\vspace{-1em}
\begin{align*}
    \mathbf{X} &= \boldsymbol{\Lambda}_1 \boldsymbol{\eta}_1 + \boldsymbol{\epsilon}_1, \\
    \boldsymbol{\eta}_1 &= \boldsymbol{\Lambda}_2 \boldsymbol{\eta}_2 + \boldsymbol{\epsilon}_2
\end{align*}

$\mathbf{X}$ are the observed variables, $\boldsymbol{\eta}_1$ are the first-order latent factors (e.g., subindices), and $\boldsymbol{\eta}_2$ is the second-order latent factor (e.g., final index). $\boldsymbol{\Lambda}_1$ and $\boldsymbol{\Lambda}_2$ are the factor loadings at each level, and $\boldsymbol{\epsilon}_1$ and $\boldsymbol{\epsilon}_2$ are the respective residual errors.

Unfortunately, this approach is impractical for small sample sizes because it must model the covariance structure between factors across levels in addition to between the items and factors. The number of parameters grows with the number of items ($p$), subindices ($q$), and their connections to one another. The errors for the traditional model can be expressed as:

\vspace{-1em}
\begin{align*}
    SE(\hat{\theta}_{\text{traditional}}) &= \sqrt{\frac{p(p+1)}{2} + \frac{q(q+1)}{2} + pq} \cdot \frac{1}{N}
\end{align*}

Standard errors therefore increase with each additional item, factor, and level of the model. Given that many international and sub-national political science indices have dozens of indicators and several subindices, traditional Confirmatory Factor Analysis will quickly run into estimation issues. As such, an alternative is needed to retain the hierarchical structure of the model while reducing the threat of parameter bias. 

As discussed earlier, we can test a novel alternative called sequential CFA, which is estimated as:

\vspace{-1em}
\begin{align*}
    \mathbf{X}^{(1)} &= \boldsymbol{\Lambda}_1 \boldsymbol{\eta}_1 + \boldsymbol{\epsilon}_1, \\
    \hat{\boldsymbol{\eta}}_1 &= \boldsymbol{\eta}_1 + \boldsymbol{\nu}_1, \\
    \mathbf{X}^{(2)} &= \boldsymbol{\Lambda}_2 \hat{\boldsymbol{\eta}}_1 + \boldsymbol{\epsilon}_2
\end{align*}

$\mathbf{X}^{(1)}$ are the observed items for the first-level CFA, and $\hat{\boldsymbol{\eta}}_1$ are the estimated factor scores from that model, which are then used as observed items in a second CFA model predicting $\mathbf{X}^{(2)}$. The term $\boldsymbol{\nu}_1$ represents factor score estimation error. Each stage is estimated independently, passing information forward sequentially. This avoids the complexity of estimating cross-level covariances while maintaining the original sample size at each step.

To understand the advantages of sequential CFA, recall the standard error complexity of traditional CFA and compare that to sequential CFA:

\vspace{-1em}
\begin{align*}
    SE(\hat{\theta}_{\text{level 1}}) &= \sqrt{\frac{p(p+1)}{2}} \cdot \frac{1}{N}, \\
    SE(\hat{\theta}_{\text{level 2}}) &= \sqrt{\frac{q(q+1)}{2}} \cdot \frac{1}{N}
\end{align*}

Because traditional CFA includes cross-level parameters, it generally results in larger standard errors. Sequential CFA reduces complexity while preserving estimation accuracy at both levels. However, sequential CFA may result in the propagation of measurement error from the first CFA model into the second. Consider the first-stage CFA:

\vspace{-1em}
\begin{align*}
    \mathbf{X} &= \boldsymbol{\Lambda}_1 \boldsymbol{\eta}_1 + \boldsymbol{\epsilon}_1, \\
    \hat{\boldsymbol{\eta}}_1 &= \boldsymbol{\eta}_1 + \boldsymbol{\nu}_1
\end{align*}

When the estimated factor scores $\hat{\boldsymbol{\eta}}_1$ are used as observed variables in the second CFA, their error affects the covariance structure:

\vspace{-1em}
\begin{align*}
    \text{Cov}(\hat{\boldsymbol{\eta}}_1) &= \boldsymbol{\Lambda}_1 \boldsymbol{\Phi}_1 \boldsymbol{\Lambda}_1' + \boldsymbol{\Psi}_{\epsilon_1} + \boldsymbol{\Psi}_{\nu_1}, \\
    \text{Cov}(\boldsymbol{\eta}_2) &= \boldsymbol{\Lambda}_2 \text{Cov}(\hat{\boldsymbol{\eta}}_1) \boldsymbol{\Lambda}_2' + \boldsymbol{\Theta}_2
\end{align*}

Error propagation therefore depends on how accurately $\hat{\boldsymbol{\eta}}_1$ represents the latent construct $\boldsymbol{\eta}_1$. Without further investigation, it is unclear how this measurement error affects the validity of the second-level factor model. This is the ultimate goal of our research, which will be investigated using a simulation study.

\section{Simulation Methodology}
We use simulations, rather than real-world data, as our primary method to investigate whether sequential CFA enables analyses of hierarchical latent factor structures with small sample sizes without introducing problematic measurement error and at what conditions these results hold. Simulations offer several advantages:

\begin{enumerate}
    \item \textbf{Effect Isolation}: Because the only difference between the two models is the use of either a sequential or traditional approach, we can make stronger claims about the performance of sequential CFA.
    \item \textbf{Power through Iterations}: Rather than relying on a few datasets, simulations allow for 100 iterations per condition, ensuring robust insights into how sequential CFA behaves.
    \item \textbf{Controlled Manipulation of Parameters}: Simulated data can be tailored to reflect a variety of realistic data-generating processes, making it possible to generalize findings across real-world studies.
\end{enumerate}

Together, these features make simulations the only effective way to evaluate the impact of sequential CFA on second-order factor score accuracy.

\subsection{Simulation Process}
To assess the comparative performance between sequential and traditional CFA, we conducted a simulation study using the \texttt{lavaan} package in R. The study generates data frames from predefined latent structures and compares the resulting factor scores produced by sequential estimation versus a full traditional CFA.

In each iteration, latent factors are simulated, observed variables are generated based on prespecified loadings, and both sequential and traditional CFA models are estimated. We then compared the resulting factor scores to the true latent values using Pearson correlations and RMSE. We repeat this process 100 times per condition, which allows for the stable estimation of performance differences.

\subsection{Models for Simulation}
Our simulation study includes three different simulation scripts for three levels of model complexity, each with a different number of latent constructs and observed indicators.

\begin{itemize}
    \item \textbf{Simple Model}: 1 Factor $\Leftarrow$ 3 Subfactors $\Leftarrow$ 6 Items
    \item \textbf{Complex Model}: 1 Factor $\Leftarrow$ 4 Subfactors $\Leftarrow$ 12 Items
    \item \textbf{Most Complex Model}: 3 Factors $\Leftarrow$ 5 Subfactors $\Leftarrow$ 15 Items (Nested Structure)
\end{itemize}

Testing across different levels of complexity is essential for understanding when sequential CFA becomes advantageous and where it may have limitations. As well, this tiered approach mirrors the diversity of real-world applications, where latent variables may be represented by simple or highly complex constructs. By incorporating a range of model sizes, our study can inform researchers about which estimation strategy is most appropriate under different empirical and theoretical constraints.

\subsection{Varying Conditions}
To further evaluate robustness, each model is simulated using a number of varying conditions. Each parameter alters the simulated data in ways that affect estimation performance: (i) Smaller sample sizes reduce statistical power, increase standard errors, and can lead to convergence failures or biased parameter estimates; (ii) Greater residual variance reduces the ability to detect meaningful patterns; (iii) Skewed distributions violate normality assumptions and can distort estimates; (iv) Heteroskedasticity violates the assumption of constant variance across indicators; and (v) Cross-loadings introduce specification errors that may affect model fit.

\vspace{1em}
\begin{table}[ht]
\centering
\captionsetup{labelformat=empty} % This hides the "Table 1" label
\caption{\textbf{Table 1: Simulation Conditions Tested Across All Models}}
\vspace{1.2em}
\renewcommand{\arraystretch}{1.4}
\setlength{\tabcolsep}{12pt}
\begin{tabular}{|c|l|}
\hline
\textbf{Parameter} & \textbf{Values Tested} \\
\hline
Sample Size & 200, 500, 1000, 2000/5000 (Not Most Complex) \\
\hline
Error Level & 0.2, 0.4, 0.6 \\
\hline
Factor Distribution & Normal, Skewed \\
\hline
Heteroskedasticity & Homoskedastic, Heteroskedastic \\
\hline
Cross-Loadings (Most Complex Only) & 0, 0.2 \\
\hline
\end{tabular}
\end{table}

\subsection{Measuring Error}
We compare the factor scores generated by CFA to the true underlying latent factors using two metrics: root mean squared error (RMSE)\footnote{As used by Zhang et al. (2024) when measuring results for an Item Response Theory study.} and Pearson correlations\footnote{As used by Kogar (2018), used in an Item Response Theory simulation study.}. These metrics are more suitable for our simulation study than traditional model fit indices\footnote{Such as the Tucker-Lewis Index, Comparative Fit Index, Root Mean Square Error of Approximation, and Standardized Root Mean Square Residual.} because they are designed for contexts where the true latent values are known. Other standard fit indices are used where true values are not known and are instead compared to other baselines (i.e., null and saturated models).

We specifically use two measures of fit to capture two different measurement dimensions. RMSE measures the average squared deviation between the estimated and true latent values. An RMSE closer to zero indicates better accuracy. Pearson correlation assesses the degree to which the estimated factor scores preserve the structure of the true latent variables. A Pearson correlation closer to one indicates a better fit.

\subsection{Simulation Results}
Each model we tested presents valuable insights into the effectiveness of sequential CFA. As such, we will detail the results of each of them individually and in turn.

\subsubsection{Simple Model}
Across the full sample, a paired \textit{t}-test revealed that sequential CFA produced significantly lower RMSEs than traditional CFA, with a mean RMSE difference of $-0.127$ at the $99.9\%$ confidence level. The $95\%$ confidence interval ranged from $-0.133$ to $-0.120$. Figure 1 (Appendix) illustrates the overall RMSE distributions for both methods. As well, Table 1 (Appendix) presents stratified mean RMSEs across all experimental conditions. Several patterns emerge:

\begin{enumerate}
    \item \textbf{Factor Distribution:} Sequential CFA consistently outperformed traditional CFA when the data followed a normal distribution across all sample sizes. In contrast, under skewed data conditions, sequential CFA underperformed relative to traditional CFA. Figure 2 (Appendix) shows density plots of RMSE distributions by method and data type, revealing that sequential CFA's RMSEs become more uniformly distributed under skewed conditions, whereas traditional CFA retains a roughly normal distribution but with higher mean RMSE.
    
    \item \textbf{Sample Size:} Assuming normality, sequential CFA outperformed traditional CFA at all sample sizes, with its robustness increasing at larger sample sizes (as is expected from any factor analysis model).

    \item \textbf{Heteroskedasticity:} Both methods demonstrated slightly improved RMSE performance under heteroskedastic conditions. This is likely due to the structured nature of the simulated error and not a reflection of sequential CFA's resilience to heteroskedasticity. In future studies, we would not incorporate heteroskedasticity.
\end{enumerate}

We also evaluated the Pearson correlation between estimated and true latent scores. A paired \textit{t}-test revealed a statistically significant difference in mean correlations, with sequential CFA exhibiting a slightly higher average correlation of $0.023$ compared to traditional CFA ($99.9\%$ confidence level; $95\%$ CI: $0.015$ to $0.031$). Figure 3 (Appendix) presents boxplots of the correlation differences (Sequential – Traditional) across all simulation conditions.

These trends are consistent with the RMSE results. Sequential CFA produced stronger correlations under normal factor distributions across all sample sizes and residual structures. However, its performance deteriorates under skewed conditions. The differences in correlation became more dispersed and even negative under some conditions. This suggests that sequential CFA is more sensitive to violations of normality in the factor distribution.

\subsubsection{Complex Model}
A paired \textit{t}-test comparing RMSEs across the full sample indicates that sequential CFA produced significantly lower RMSEs than traditional CFA, with a mean difference of $-0.247$ at the $99.9\%$ confidence level. The $95\%$ confidence interval ranged from $-0.254$ to $-0.239$. Figure 4  displays a boxplot of RMSE differences between the two methods.

As with the simple model, condition-specific results provide necessary context. Table 2 (Appendix) presents stratified RMSE means by sample size, data type, and residual pattern. Appendix 5 (Appendix) also shows the RMSE density distribution by method and data type. We notice all major themes from the simple model re-emerging with our simulations of the complex model. 

The comparison of Pearson correlations is also consistent with the findings of the simple model. A paired \textit{t}-test found that sequential CFA significantly outperformed traditional CFA, with a mean difference of $0.076$ at the $99.9\%$ confidence level. The $95\%$ confidence interval ranged from $0.068$ to $0.084$. Figure 6 (Appendix) provides a boxplot comparing mean Pearson correlations across all conditions. These trends mirrored the RMSE results, with sequential CFA producing higher correlations under normal data distributions, regardless of residual structure or sample size. However, under skewed distributions, the differences in correlation become more spread out, less consistent, and at times negative.

\subsubsection{Most Complex Model}
The most complex model consists of three factors, five subfactors, and fifteen items. However, the model incorporates a nested design for added complexity. The items are first grouped into five subfactors, each measured by three items (e.g., $F_1 =~ V1 + V2 + V3$). These subfactors are hierarchically structured: $G_1$ loads on $F_1$ and $F_2$, $G_2$ on $F_3$ and $F_4$, and the highest-order factor $H$ loads on $G_1$, $G_2$, and $F_5$.

A paired \textit{t}-test comparing RMSEs across the full sample reveals a notable shift. Unlike previous models, traditional CFA marginally outperformed sequential CFA, with a mean RMSE difference of $0.015$ at the $99.9\%$ confidence level (95\% CI: $0.012$ to $0.017$). Figure 7 (Appendix) presents a boxplot of these RMSE differences.

When stratifying results by sample size, distribution, and residual structure (Table 3 of the Appendix), this pattern holds consistently across all conditions. Figure 8 (Appendix) displays the RMSE distributions by method and data type. In contrast to earlier models, the density curves for sequential and traditional CFA are nearly indistinguishable across all data conditions, which indicates performance convergence as model complexity increases.

Pearson correlation results also reflect this reversal. A paired \textit{t}-test indicates that traditional CFA outperformed sequential CFA by a small margin. The mean correlation difference was $-0.0098$ at the $99.9\%$ level (95\% CI: $-0.013$ to $-0.007$). Figure 9 (Appendix) provides a boxplot comparison of correlation estimates across all conditions. Ultimately, the results collectively indicate that the two methods perform similarly under complex conditions.

\section{External Validation}
In addition to the above simulation study, we also apply sequential CFA to an existing index (the World Justice Project's (WJP) "Rule of Law Index" (ROL)) to validate the simulation results. 

\subsection{The Rule of Law Index}
The WJP has published an annual "Rule of Law Index" since 2008. The Rule of Law Index covers between 100 and 125 countries depending on the year. The ROL Index measures each participating country's "durable system of laws, institutions, norms, and community commitment that delivers four universal principles: accountability, just law, open government, and accessible and impartial justice" (World Justice Project, 2025). 

The ROL Index is hierarchical and made up of over 500 variables, 44 sub-factors, eight factors, and one overall index. The 500+ variables are derived from nationally representative household surveys and expert questionnaires, capturing perceptions and experiences related to justice, corruption, transparency, and institutional effectiveness. These variables are grouped into 44 sub-factors. Each sub-factor is constructed by aggregating the underlying question-level variables into a normalized score ranging from 0 to 1. These items are normalized and then averaged within each sub-factor to ensure comparability across countries. No single question is weighted more heavily than others within a sub-factor unless explicitly noted.

These sub-factors in turn contribute to eight factors: Constraints on Government Powers, Absence of Corruption, Open Government, Fundamental Rights, Order and Security, Regulatory Enforcement, Civil Justice, and Criminal Justice. Each factor represents a distinct aspect of the rule of law, and these are also scored on a 0 to 1 scale.

Finally, the overall Rule of Law Index score for each country is calculated as the arithmetic mean of the eight factor scores. This score provides a summary measure of a country's overall adherence to the rule of law.

\subsection{Validation Process}
We apply both traditional and sequential CFA to estimate the factors and the overall index.\footnote{Sub-factor, factor, and overall index data is made available by WJP for each participating country. However, WJP does not make the household-level data available, and so it is only possible to estimate the Rule of Law Index at the factor and index levels.} We collected the raw data directly from WJP for all participating countries and years between 2012 to 2024 and performed CFA for each year independently.

Before reviewing the result, we check to ensure that the CFA estimator has converged and produced valid loadings. If the model succeeds, then various tests of fit and other diagnostics are evaluated. Specifically, we examine four commonly used global fit indices: the Comparative Fit Index (CFI), the Tucker–Lewis Index (TLI), the Root Mean Square Error of Approximation (RMSEA), and the Standardized Root Mean Square Residual (SRMR). In addition to overall model fit, we report McDonald's omega to assess internal consistency reliability, as it provides a more accurate estimate than Cronbach's alpha when factor loadings are unequal. These diagnostics help determine whether the latent factor structure is empirically supported and whether the items coherently measure the intended construct.

If the model fails, we apply minor adjustments to ensure that the model could fit under different conditions. If the model continues to fail, then we conclude that the model is unsuitable for the underlying data.

\subsection{Validation Results}
We first attempted to model the ROL Index using traditional CFA. The model consistently fails to provide valid results and returns "NA" for each standardized loading. This indicates that the CFA is unable to model the provided data given its structure. We then relaxed the assumption that each item loads on only one factor (although even this violates the assumptions of measurement provided by WJP). However, even with this adjustment, the model continued to produce inadmissible estimates. These issues suggest that the factor structure may not be well-defined or that the data may not support a traditional CFA approach. Similarly, we also attempted to model the ROL Index using Bayesian CFA, which is frequently relied upon by social scientists where there are small samples. However, Bayesian CFA also failed and was unable to produce valid posteriors despite adjusting priors. This is the product of both the complexity of the ROL Index and also the small sample size.

In contrast, the sequential CFA model instantly provided valid results. We assessed model fit and reliability across both stages of the model hierarchy. The first-stage CFA generally shows strong model performance. McDonald's omega scores consistently show that the model is internally reliable. Most year averages consistently exceed the 0.70 threshold. Global fit indices for the first-stage CFAs support the reliability findings. The CFI and SRMR are generally within or around ideal ranges (CFI~$>~0.9$, SRMR~$<~0.08$) across years, with some variation depending on the factor. While the RMSEA values in this analysis frequently exceed conventional thresholds for acceptable fit, this pattern is not unexpected given the small number of indicators per factor in several first-stage models. RMSEA is known to over-penalize models with low degrees of freedom, which is typical in narrowly specified CFA models (Kenny, Kaniskan and McCoach, 2014). This effect is even more pronounced in small samples. As well, since sequential CFA is most likely to be applied for small samples, these RMSEA scores are unconcerning. Rather, more weight should be given to indices like CFI, TLI, and SRMR in evaluating model fit in such contexts. Table 2 below details the results of the first-stage sequential CFA averaging the collective tests of fit and omegas for each year:

\begin{table}[ht]
\centering
\caption{First-Stage CFA Fit Indices and Reliability (All Years, Averaged)}
\label{tab:first_stage_cfa_all}
\begin{tabular}{|l|c|c|c|c|c|}
\hline
	extbf{Year} & \textbf{CFI} & \textbf{TLI} & \textbf{RMSEA} & \textbf{SRMR} & \textbf{Omega} \\
\hline
2012--2013 & 0.879 & 0.806 & 0.238 & 0.047 & 0.663 \\
2014      & 0.889 & 0.811 & 0.257 & 0.044 & 0.691 \\
2015      & 0.894 & 0.813 & 0.246 & 0.040 & 0.709 \\
2016      & 0.902 & 0.823 & 0.254 & 0.042 & 0.697 \\
2017--2018 & 0.900 & 0.825 & 0.248 & 0.042 & 0.705 \\
2019      & 0.897 & 0.811 & 0.269 & 0.042 & 0.713 \\
2020      & 0.897 & 0.811 & 0.271 & 0.042 & 0.718 \\
2021      & 0.905 & 0.824 & 0.267 & 0.039 & 0.729 \\
2022      & 0.910 & 0.835 & 0.260 & 0.037 & 0.743 \\
2023      & 0.913 & 0.841 & 0.256 & 0.036 & 0.747 \\
2024      & 0.912 & 0.838 & 0.262 & 0.036 & 0.753 \\
\hline
\end{tabular}
\end{table}

The second-stage CFA, which models a single latent factor from the eight subindex scores, also demonstrates strong validity. Reliability remains consistently high, with omega values ranging from 0.78 to 0.82 across all years. Fit indices decline modestly over time, with CFI and TLI dropping from above 0.90 in earlier years to around 0.84–0.78 in later years. Most CFI and all SRMR values are in ideal or acceptable ranges. RMSEA values are still high but should be interpreted with caution. Below at Table 3 are the results of the second-stage CFA.

\begin{table}[ht]
\centering
\caption{Second-Stage CFA Fit Indices and Reliability (All Years)}
\label{tab:second_stage_cfa_all}
\begin{tabular}{|l|c|c|c|c|c|}
\hline
\textbf{Year} & \textbf{CFI} & \textbf{TLI} & \textbf{RMSEA} & \textbf{SRMR} & \textbf{Omega} \\
\hline
2012--2013 & 0.927 & 0.898 & 0.198 & 0.036 & 0.779 \\
2014      & 0.946 & 0.925 & 0.176 & 0.028 & 0.793 \\
2015      & 0.891 & 0.847 & 0.268 & 0.043 & 0.822 \\
2016      & 0.879 & 0.831 & 0.271 & 0.044 & 0.807 \\
2017--2018 & 0.869 & 0.817 & 0.281 & 0.048 & 0.797 \\
2019      & 0.863 & 0.808 & 0.289 & 0.051 & 0.787 \\
2020      & 0.854 & 0.795 & 0.300 & 0.051 & 0.783 \\
2021      & 0.851 & 0.792 & 0.310 & 0.048 & 0.802 \\
2022      & 0.854 & 0.796 & 0.308 & 0.046 & 0.800 \\
2023      & 0.842 & 0.778 & 0.322 & 0.048 & 0.796 \\
2024      & 0.845 & 0.782 & 0.320 & 0.047 & 0.795 \\
\hline
\end{tabular}
\end{table}

Taken together, the results indicate that the measurement model is relatively reliable and structurally sound. Despite the increased complexity of the model over time, the sequential CFA approach yields interpretable and valid factor scores that support the construction of a robust Rule of Law Index that is arguably more representative of the underlying latent construct WJP is measuring.

\subsection{Illustrative Examples}
I compare individual country factor score results generated by sequential CFA and the WJP ROL Score to isolate divergent responses to observable events. We focus on two substantial events in global politics: The Thailand Coup of 2014 and the Sudanese political unrest beginning around the end of 2021 and beginning of 2022. A time series comparison for Thailand and Sudan of sequential CFA factor scores and the WJP ROL scores are located at Figure 10.

There is a visible difference between the sequential CFA results and the WJP ROL Score beginning in 2014. The WJP ROL Score registers a slight decline in the Rule of Law starting in 2014, but these changes are minimal throughout the sample years. In contrast, the sequential CFA factor scores record an immediate and substantial drop beginning in 2014. There is a quick recovery in perceptions of the Rule of Law, but this is consistent with a potential stability achieved by military control immediately following a political crisis. As well, sequential CFA records another dramatic decline in 2017, which coincides with the year the military regime replaced Thailand's constitution to entrench military control and restrict the power of elected officials.

Additionally, despite coups and widespread ethnic violence, the WJP ROL Scores only reveal a marginal reduction beginning in 2022. The sequential CFA factor scores follow a sharp decrease beginning in 2022 that far outpace the WJP ROL Scores.

These two case examples highlight the primary issue with average weighting. All components of an index are rarely equally important, especially where their scores are meant to also register material global issues. Sequential CFA, as a data-driven method, consistently captures important events because it measures the underlying latent construct, which is dependent on different factors that have the freedom to change over a time series.

\section{Discussion}
The results from our simulation study show an encouraging picture for using sequential CFA in small-sample hierarchical latent variable modeling. Across the simple and complex model structures, sequential CFA consistently produced lower RMSEs and higher Pearson correlations when compared to traditional CFA. Sequential CFA instead performed better because of its staggered design, which reduced the number of parameters to be estimated at each stage and in turn provide more accurate estimates. This suggests that sequential CFA may provide a reliable estimation method with a lower sample-size threshold for researchers working with limited observations.

However, the results from the skewed data simulations clarify the overall findings. For all three models, skewed factor distributions led to markedly higher RMSEs for sequential CFA with uniform-like distributions. This pattern signals the breakdown in the statistical assumptions that ensure reliable parameter estimation in CFA and affirms our initial hypothesis regarding error propagation. When first-stage factor scores in a sequential CFA fail to capture the true distribution of the latent variables, the second-stage model reinforces these initial distortions rather than mitigating them. In contrast, traditional CFA employs simultaneous estimation which enables partial correction for measurement error throughout the model. Sequential CFA cannot adjust for this propagated bias because the error-filled first-stage factor scores are already treated as observables by the second-stage. While traditional CFA already includes normality as a statistical assumption, this factor becomes even more important for sequential CFA.

Our application of sequential CFA to the ROL Index reveals the core strength of this novel method. Capturing and measuring the underlying latent factors behind observed data requires significantly more precision and must meet more structural demands than averaged or arbitrary weights. We observe through our study that traditional frequentist or Bayesian latent factor analysis on the ROL Index is impossible. However, sequential Confirmatory Factor Analysis has reduced enough model parameters to allow for valid estimated results. Additionally, tests of fit and McDonald's omega values at the first and second stages of sequential CFA reaffirm that the factor-level and index-level scores accurately represent the underlying latent constructs that form WJP's ROL Index.

Generally, our study offers a valuable contribution to measurement research in political science. Sequential CFA offers a viable method to measure latent structures when full model estimation is otherwise infeasible due to sample size. Further, in low or moderate complexity models with relatively normal data, sequential CFA consistently outperforms traditional CFA. Considering that this method is restricted to hierarchical models with limited complexity, this contribution may not seem revolutionary on its face. However, sample size constraints frequently eliminate useful research ideas. Sequential CFA, if available as an alternative to other inaccessible measurement methods, may enable the construction of indices that otherwise may not have been possible beforehand, either through traditional CFA or Bayesian CFA. This is especially important for researchers conducting cross-national or sub-national analyses where the number of observations is fixed.

\section{Limitations}
There are several limitations to our study. These limitations do not undermine the validity of the findings where they apply, but instead highlight how generalizable the results are. As well, these limitations also point to directions for future methodological development for this method and others that may deconstruct hierarchical measurement models.

First, these results are specific to CFA and do not implicitly extend to other common measurement modeling approaches, such as MCMC (which handles estimation and factor scoring differently). Structural and algorithmic differences across methods mean that the staggered estimation logic used in sequential CFA should not be transferred to other techniques without a separate and tailored simulation study. Therefore, the conclusions drawn in our study are bound to CFA alone.

Second, our simulation study design cannot fully replicate all possible combinations of varying data conditions in the real world. Political science data often includes complications such as clusters, autocorrelation, and heavy-tailed distributions, among other irregularities. Although our simulations reflect plausible data, it is possible that non-tested forms of data irregularity could impact how well sequential CFA performs. With more time and resources, we would create many more models to simulate with more complex already-existing features (i.e., more diverse and applied cross-loadings) in addition to features not tested at all (i.e., autocorrelation). Researchers applying this method to empirical datasets should therefore conduct rigorous data diagnostic checks and use sequential CFA cautiously.

Third, the sequential CFA strategy is only appropriate for hierarchical CFA models. These models, while useful for indices, are still a minority of CFA applications. Many researchers use single-factor models where the sequential estimation framework is inapplicable. As such, while our study contributes a valuable methodological option, it is not a universal solution to the problem of small-N estimation in CFA.

Fourth, the advantages of sequential CFA are contingent on the normality of the underlying data, particularly at the first stage of estimation. In our simulations, this manifests as error propagation into the second-stage model. However, this limitation is not unique to sequential CFA and reflects the underlying assumption of multivariate normality in all CFA models. Traditional CFA also relies on approximately normal latent distributions, and its performance deteriorates under severe skewness, though simultaneous estimation can partially absorb some error. Researchers applying sequential CFA to skewed data should therefore consider standard robustness strategies, such as applying Box–Cox or Yeo–Johnson transformations to first-stage factors or winsorizing extreme items. In practice, these strategies align sequential CFA with the same precautions required for any CFA-based index construction.

\section{Conclusion}
Political science researchers often struggle with constructing indices using CFA because of sample size constraints. Typically, hundreds or thousands of observations are required to conduct even a simple CFA, which in turn prevents many political scientists from conducting national-level observational studies. 

A potential solution to sample size restrictions in hierarchical measurement models is sequential CFA, which estimates two or more levels of factors separately from one another. By treating factor scores as observable variables in the sequential CFA estimation, you implicitly retain your sample size while removing cross-level covariance estimation. The primary barrier to using sequential CFA is the potential for errors in the first-step CFA propagating into further iterations.

Our research demonstrates that sequential CFA is a viable alternative to traditional traditional CFA under certain circumstances. Sequential CFA outperforms traditional CFA across most model complexities, sample sizes, and covariance matrix conditions. Sequential CFA is even robust to data with heteroskedasticity. Most importantly, sequential CFA can provide valid estimates in many circumstances where traditional CFA cannot be modeled at all. However, sequential CFA generally under-performs against traditional CFA where the data is materially skewed. The likely reason for this is that CFA generally assumes normality, and thus the bias from non-normal data in the first CFA will propagate into the second CFA if the first level factors are observable rather than latent.

While our proposed approach is methodologically straightforward, its implications are significant. This strategy enables latent measurement modeling in empirical contexts where traditional CFA frameworks often fail, offering researchers a reliable and interpretable path to construct hierarchical latent indices.

% --------------------------
% --------------------------
\clearpage
\nocite{*}

% --------------------------
% Bibliography (inserted directly for arXiv)
% --------------------------

% --------------------------
% Appendix with Figures and Tables
% --------------------------

\clearpage
\appendix
\section*{Appendix: Figures and Tables}
\setcounter{table}{0}

% SIMPLE MODEL FIGURES/TABLES
\begin{figure}[htbp]
  \centering
  \includegraphics[width=0.7\textwidth]{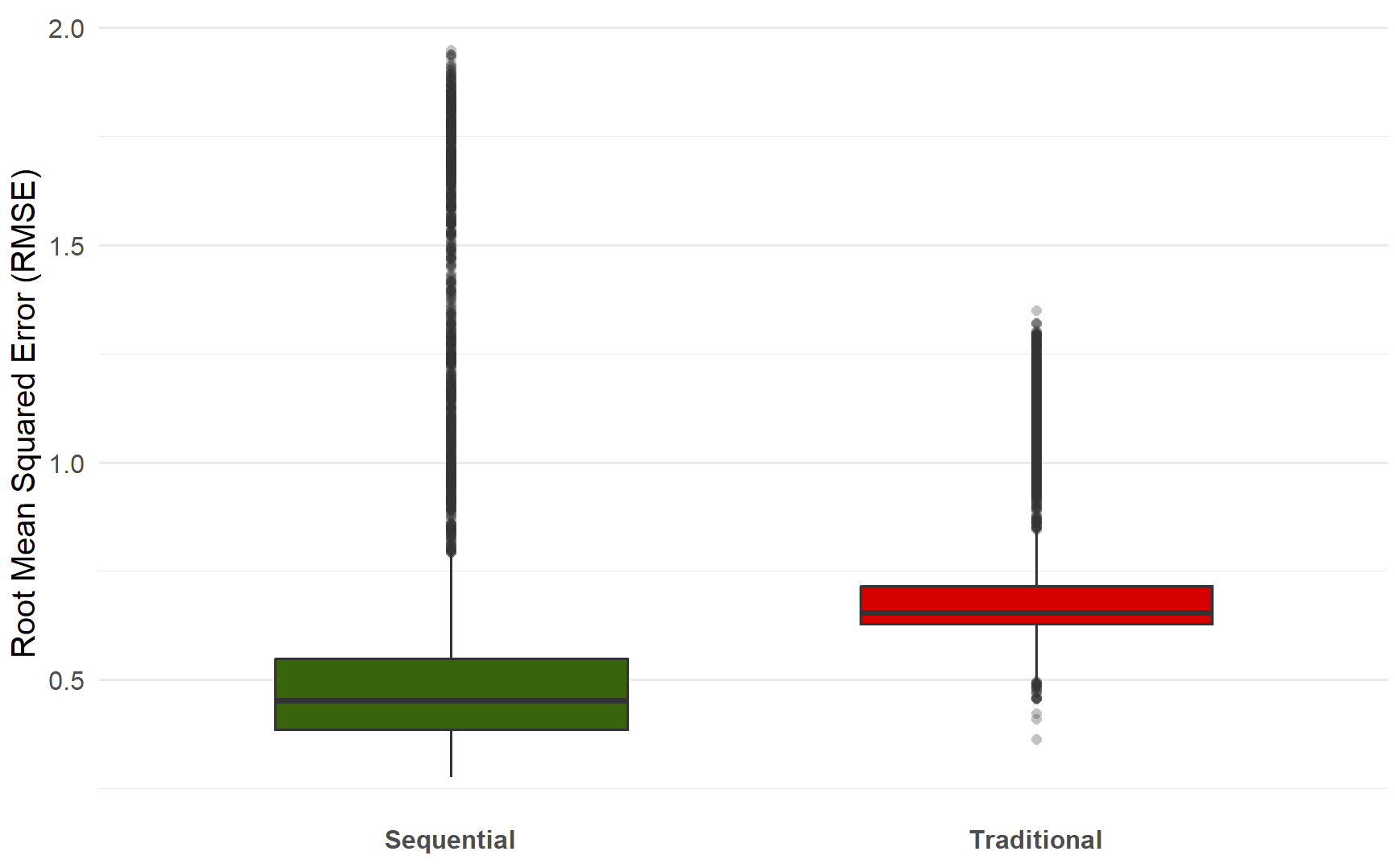}
  \caption{Simple Model: RMSE Distributions Across Methods}
  \label{fig:simple_rmse_boxplot}
\end{figure}

\begin{table}[ht]
\centering
\caption{Simple Model - Mean RMSE by Sample Size, Factor Distribution, and Heteroskedasticity}
\label{tab:rmse_summary}
\begin{tabular}{|r|l|l|r|r|}
\hline
\textbf{n} & \textbf{Distribution} & \textbf{Residual Pattern} & \textbf{Sequential RMSE} & \textbf{Traditional RMSE} \\
\hline
100  & Normal & Heteroskedastic   & 0.560 & 0.678 \\
100  & Normal & Homoskedastic     & 0.563 & 0.685 \\
100  & Skewed & Heteroskedastic   & 1.367 & 1.081 \\
100  & Skewed & Homoskedastic     & 1.357 & 1.081 \\
\hline
500  & Normal & Heteroskedastic   & 0.428 & 0.649 \\
500  & Normal & Homoskedastic     & 0.428 & 0.648 \\
500  & Skewed & Heteroskedastic   & 1.375 & 1.096 \\
500  & Skewed & Homoskedastic     & 1.288 & 1.063 \\
\hline
1000 & Normal & Heteroskedastic   & 0.414 & 0.646 \\
1000 & Normal & Homoskedastic     & 0.415 & 0.644 \\
1000 & Skewed & Heteroskedastic   & 1.397 & 1.085 \\
1000 & Skewed & Homoskedastic     & 1.305 & 1.086 \\
\hline
2000 & Normal & Heteroskedastic   & 0.408 & 0.644 \\
2000 & Normal & Homoskedastic     & 0.409 & 0.644 \\
2000 & Skewed & Heteroskedastic   & 1.305 & 1.076 \\
2000 & Skewed & Homoskedastic     & 1.353 & 1.094 \\
\hline
5000 & Normal & Heteroskedastic   & 0.404 & 0.643 \\
5000 & Normal & Homoskedastic     & 0.405 & 0.645 \\
5000 & Skewed & Heteroskedastic   & 1.338 & 1.086 \\
5000 & Skewed & Homoskedastic     & 1.411 & 1.096 \\
\hline
\end{tabular}
\end{table}

\begin{figure}[htbp]
  \centering
  \includegraphics[width=0.8\textwidth]{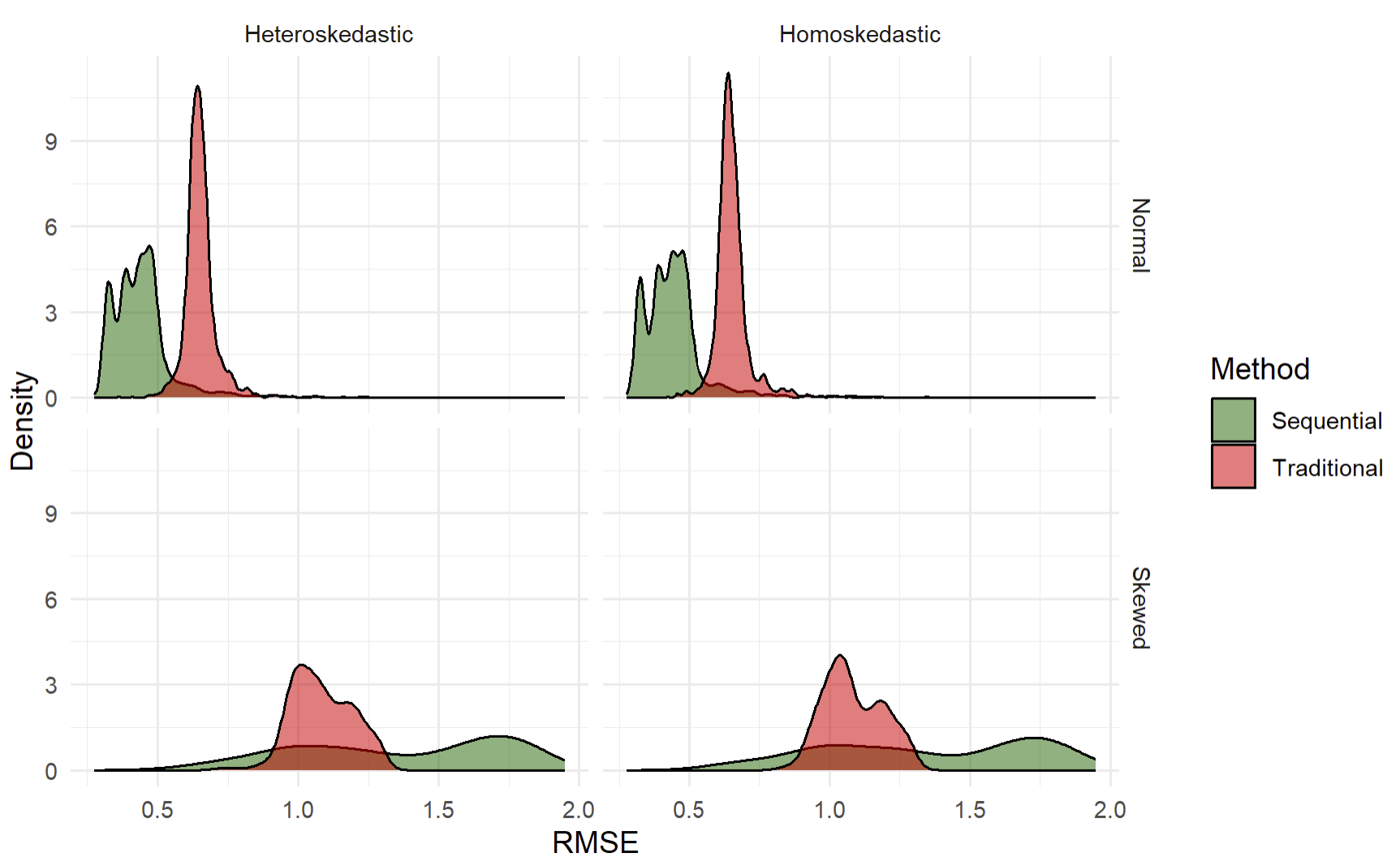}
  \caption{Simple Model - RMSE Density Distributions by Method and Data Type}
  \label{fig:simple_rmse_density}
\end{figure}

\begin{figure}[htbp]
  \centering
  \includegraphics[width=0.8\textwidth]{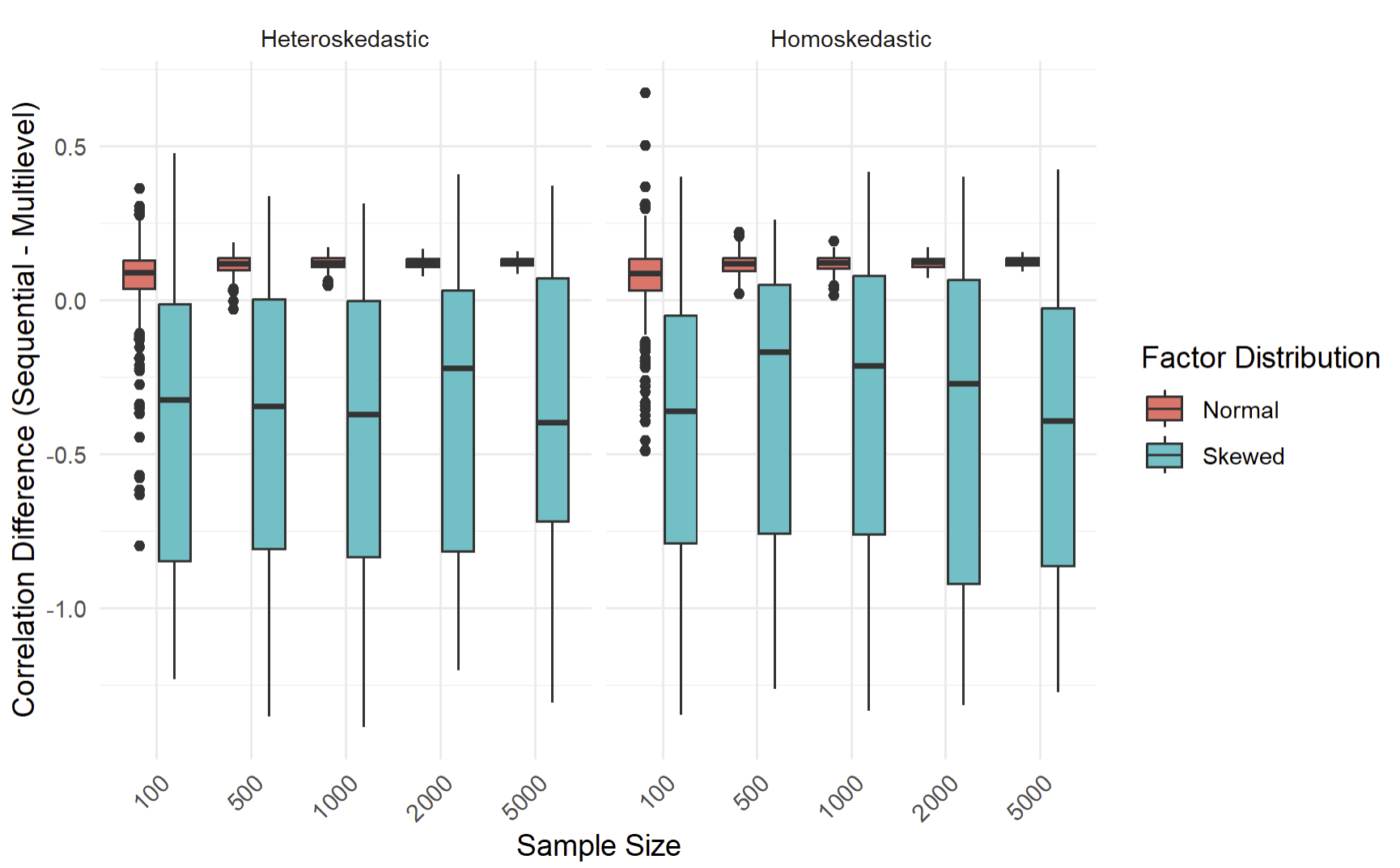}
  \caption{Simple Model - Pearson Correlation Differences Across Conditions}
  \label{fig:simple_pcor_boxplot}
\end{figure}

% COMPLEX MODEL FIGURES/TABLES
\begin{figure}[htbp]
  \centering
  \includegraphics[width=0.7\textwidth]{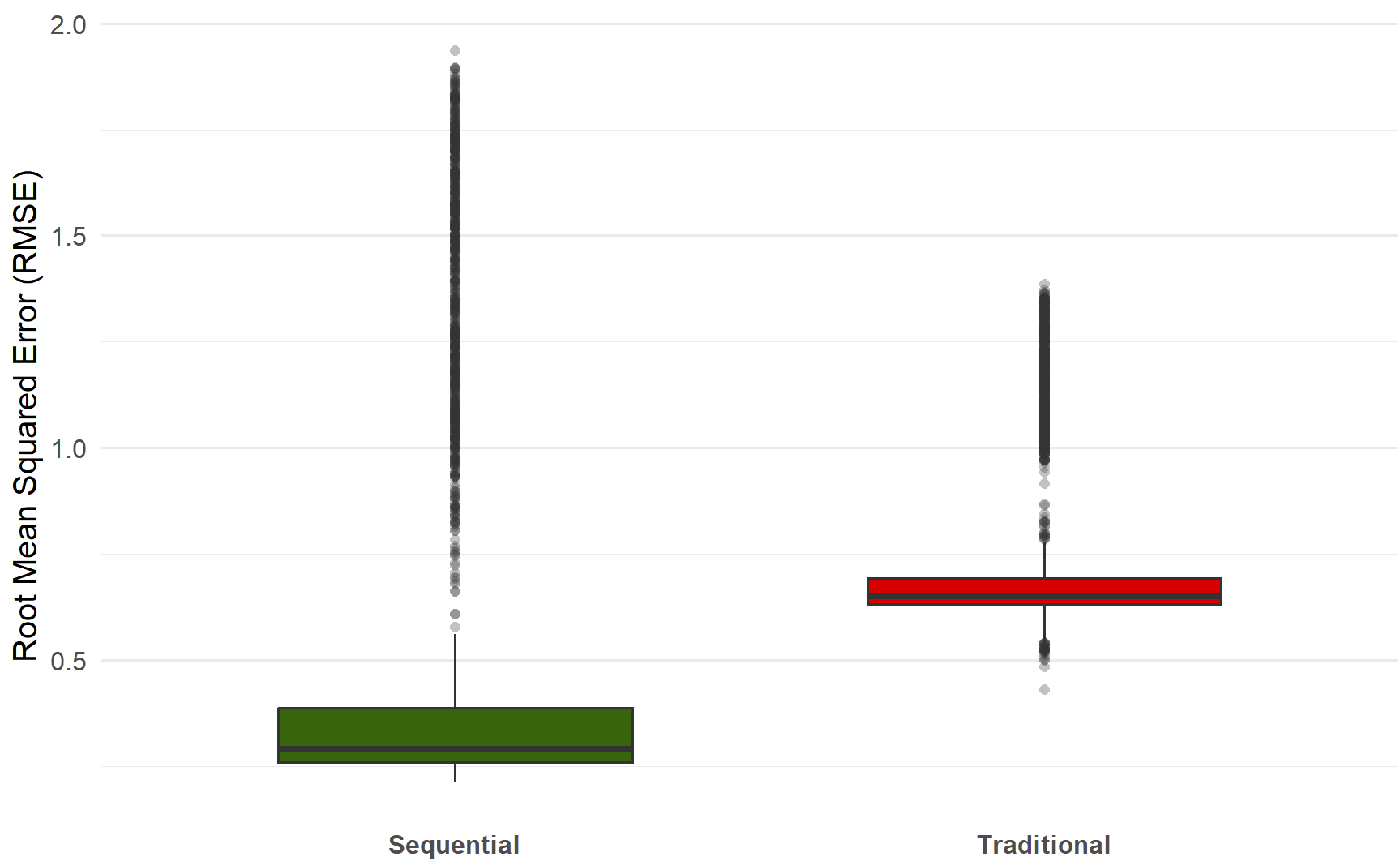}
  \caption{Complex Model - RMSE Distributions Across Methods}
  \label{fig:complex_rmse_boxplot}
\end{figure}

\begin{table}[ht]
\centering
\caption{Complex Model - Mean RMSE by Sample Size, Factor Distribution, and Heteroskedasticity}
\label{tab:rmse_summary_bordered}
\begin{tabular}{|r|l|l|r|r|}
\hline
\textbf{n} & \textbf{Distribution} & \textbf{Residual Pattern} & \textbf{Sequential RMSE} & \textbf{Traditional RMSE} \\
\hline
100 & Normal & Heteroskedastic & 0.332 & 0.652 \\
100 & Normal & Homoskedastic & 0.371 & 0.661 \\
100 & Skewed & Heteroskedastic & 1.249 & 1.104 \\
100 & Skewed & Homoskedastic & 1.368 & 1.147 \\
\hline
500 & Normal & Heteroskedastic & 0.272 & 0.641 \\
500 & Normal & Homoskedastic & 0.307 & 0.648 \\
500 & Skewed & Heteroskedastic & 1.34 & 1.122 \\
500 & Skewed & Homoskedastic & 1.391 & 1.139 \\
\hline
1000 & Normal & Heteroskedastic & 0.263 & 0.639 \\
1000 & Normal & Homoskedastic & 0.299 & 0.642 \\
1000 & Skewed & Heteroskedastic & 1.376 & 1.119 \\
1000 & Skewed & Homoskedastic & 1.363 & 1.134 \\
\hline
2000 & Normal & Heteroskedastic & 0.26 & 0.641 \\
2000 & Normal & Homoskedastic & 0.296 & 0.642 \\
2000 & Skewed & Heteroskedastic & 1.377 & 1.152 \\
2000 & Skewed & Homoskedastic & 1.385 & 1.172 \\
\hline
5000 & Normal & Heteroskedastic & 0.258 & 0.641 \\
5000 & Normal & Homoskedastic & 0.293 & 0.642 \\
5000 & Skewed & Heteroskedastic & 1.427 & 1.161 \\
5000 & Skewed & Homoskedastic & 1.384 & 1.163 \\
\hline
\end{tabular}
\end{table}

\begin{figure}[htbp]
  \centering
  \includegraphics[width=0.8\textwidth]{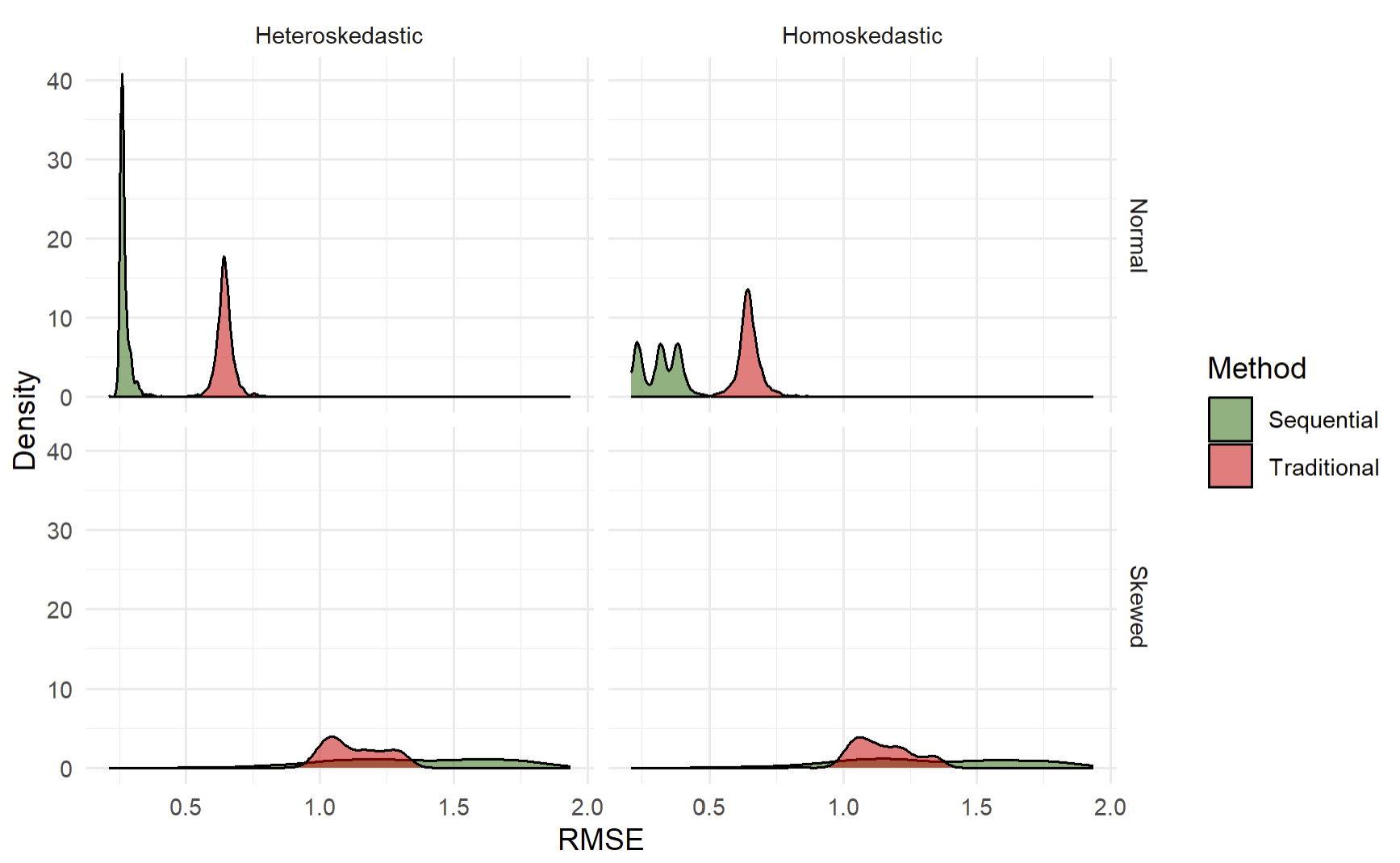}
  \caption{Complex Model - RMSE Density Distributions by Method and Data Type}
  \label{fig:complex_rmse_density}
\end{figure}

\begin{figure}[htbp]
  \centering
  \includegraphics[width=0.8\textwidth]{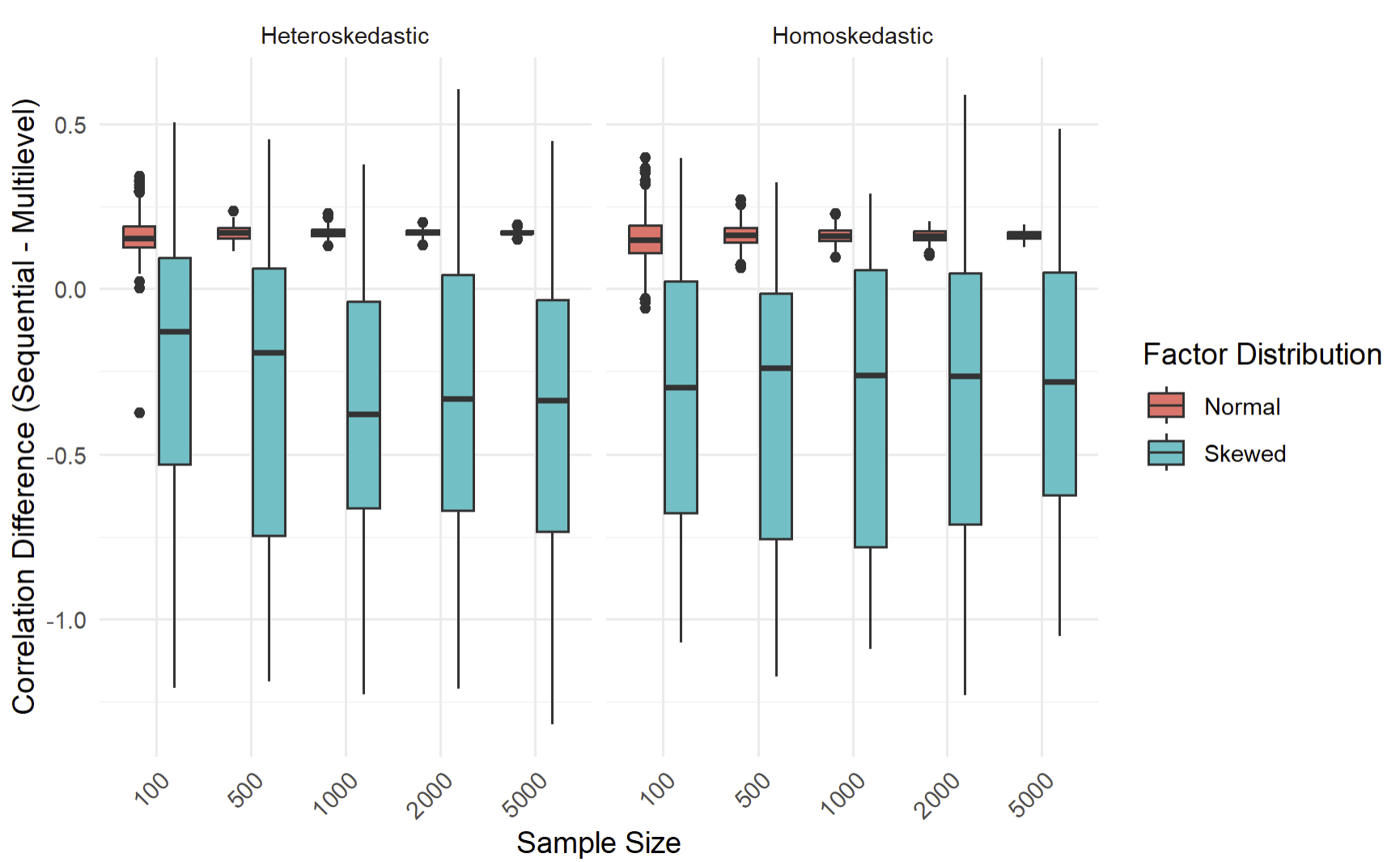}
  \caption{Complex Model - Pearson Correlation Differences Across Conditions}
  \label{fig:complex_pcor_boxplot}
\end{figure}

% MOST COMPLEX MODEL FIGURES/TABLES
\begin{figure}[htbp]
  \centering
  \includegraphics[width=0.7\textwidth]{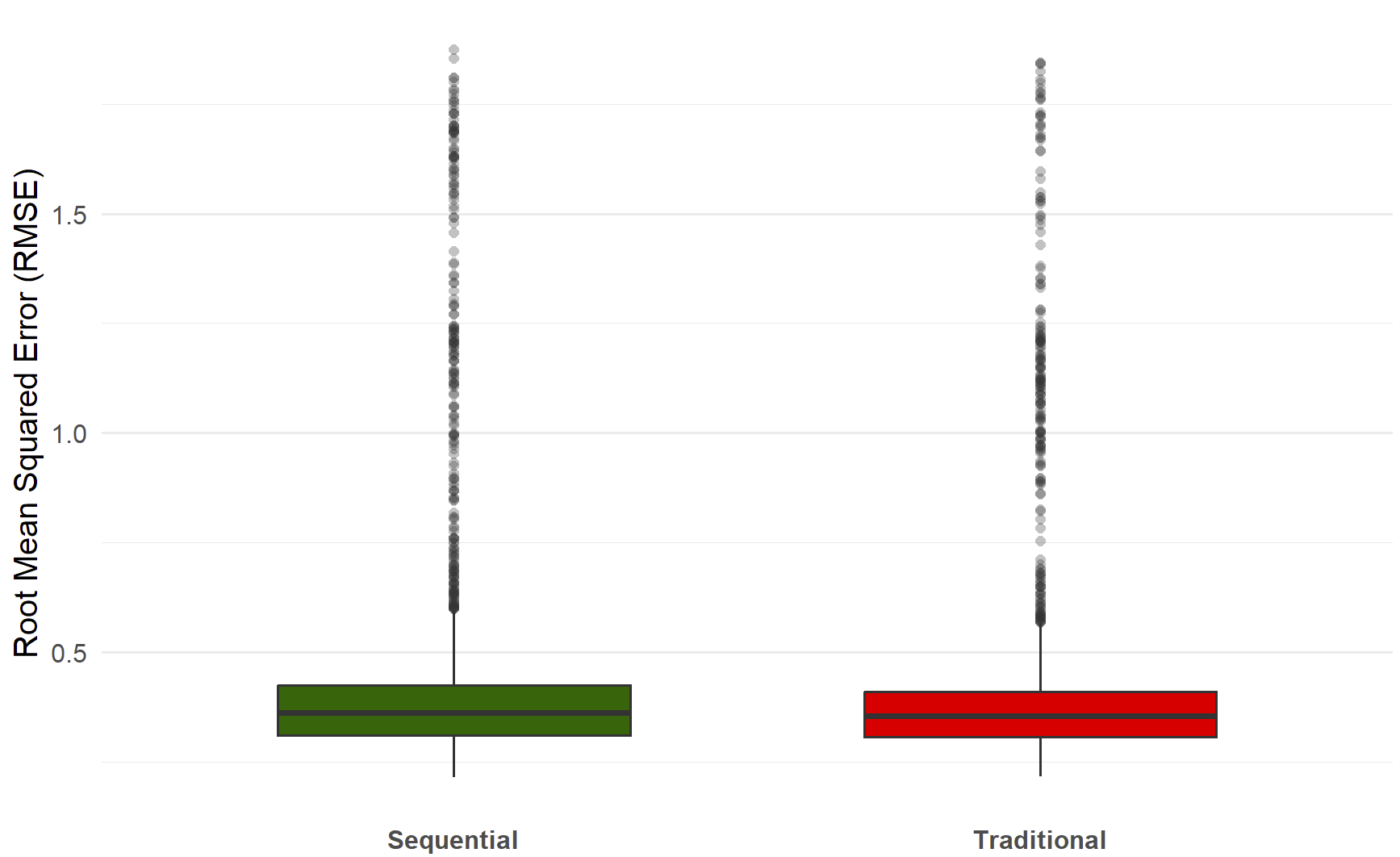}
  \caption{Most Complex Model - RMSE Distributions Across Methods}
  \label{fig:mostcomplex_rmse_boxplot}
\end{figure}

\begin{table}[ht]
\centering
\caption{Most Complex Model - Mean RMSE by Sample Size, Factor Distribution, and Heteroskedasticity}
\label{tab:rmse_complex}
\begin{tabular}{|r|l|l|r|r|}
\hline
\textbf{n} & \textbf{Distribution} & \textbf{Residual Pattern} & \textbf{Sequential RMSE} & \textbf{Traditional RMSE} \\
\hline
200  & Normal & Heteroskedastic   & 0.399 & 0.381 \\
200  & Normal & Homoskedastic     & 0.441 & 0.413 \\
200  & Skewed & Heteroskedastic   & 1.235 & 1.207 \\
200  & Skewed & Homoskedastic     & 1.387 & 1.234 \\
\hline
500  & Normal & Heteroskedastic   & 0.343 & 0.336 \\
500  & Normal & Homoskedastic     & 0.371 & 0.362 \\
500  & Skewed & Heteroskedastic   & 1.367 & 1.307 \\
500  & Skewed & Homoskedastic     & 1.318 & 1.155 \\
\hline
1000 & Normal & Heteroskedastic   & 0.327 & 0.323 \\
1000 & Normal & Homoskedastic     & 0.352 & 0.347 \\
1000 & Skewed & Heteroskedastic   & 1.330 & 1.269 \\
1000 & Skewed & Homoskedastic     & 1.325 & 1.231 \\
\hline
\end{tabular}
\end{table}

\begin{figure}[htbp]
  \centering
  \includegraphics[width=0.8\textwidth]{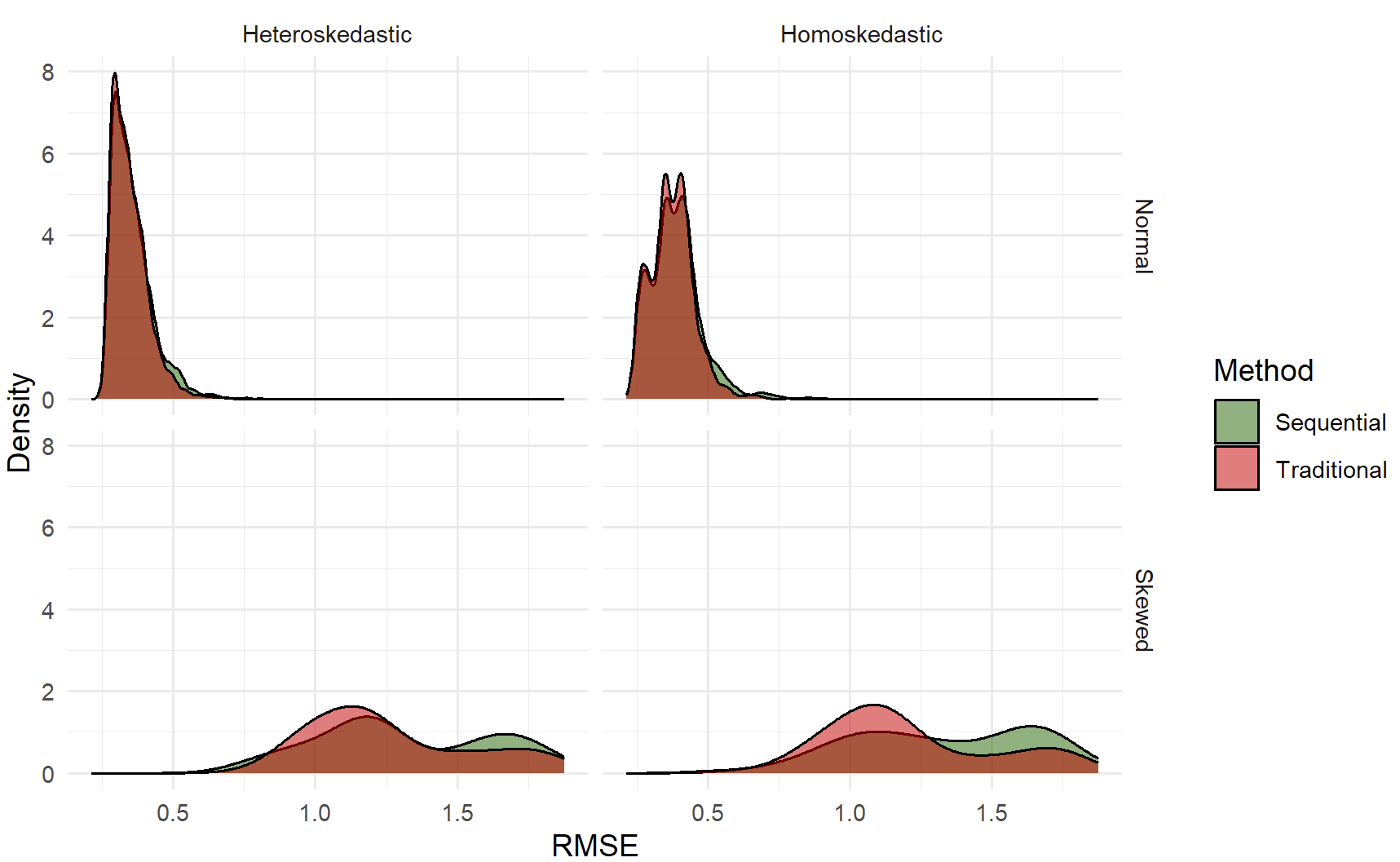}
  \caption{Most Complex Model - RMSE Density Distributions by Method and Data Type}
  \label{fig:mostcomplex_rmse_density}
\end{figure}

\begin{figure}[htbp]
  \centering
  \includegraphics[width=0.8\textwidth]{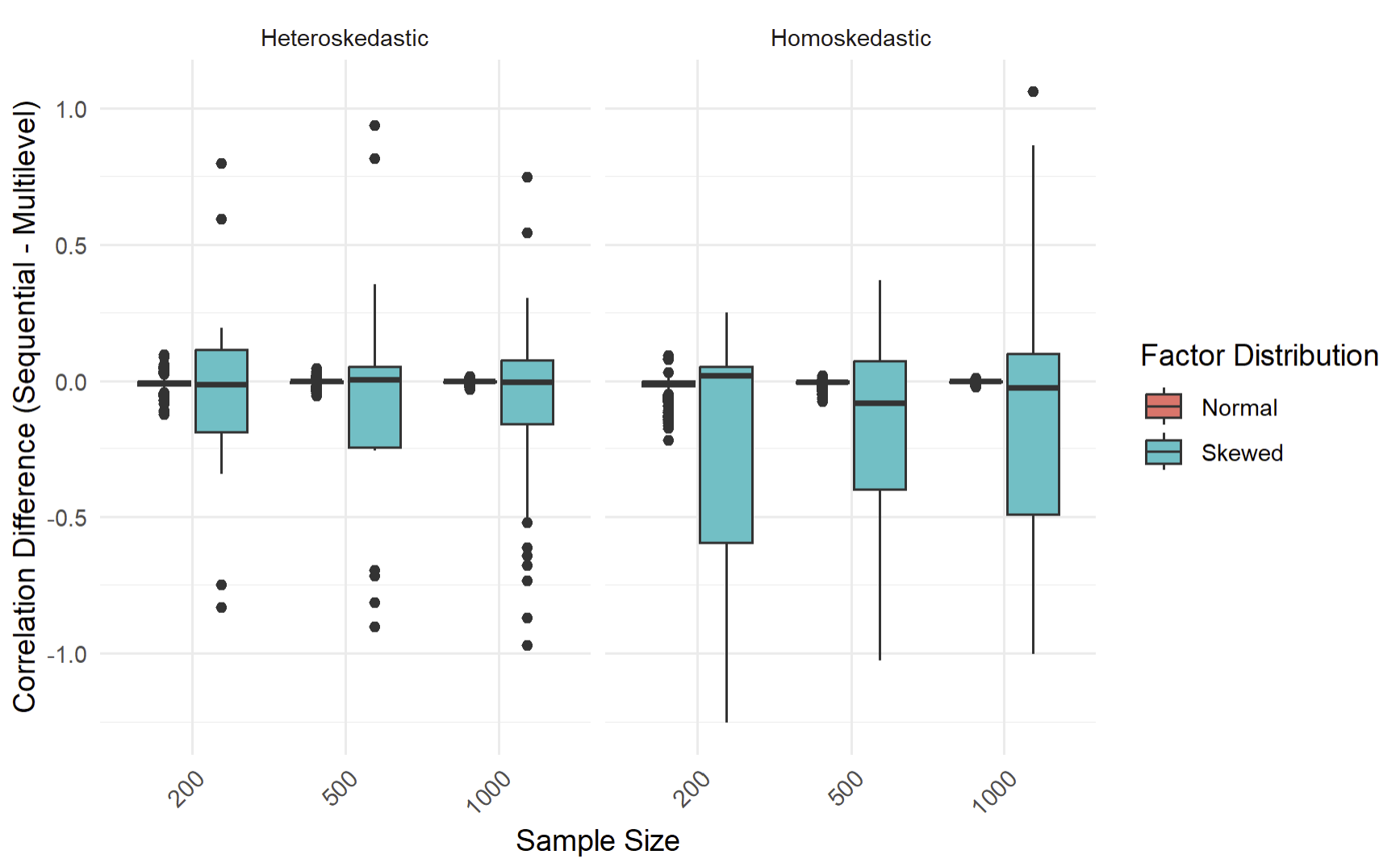}
  \caption{Most Complex Model - Pearson Correlation Differences Across Conditions}
  \label{fig:mostcomplex_pcor_boxplot}
\end{figure}

\begin{figure}[htbp]
  \centering
  \includegraphics[width=0.8\textwidth]{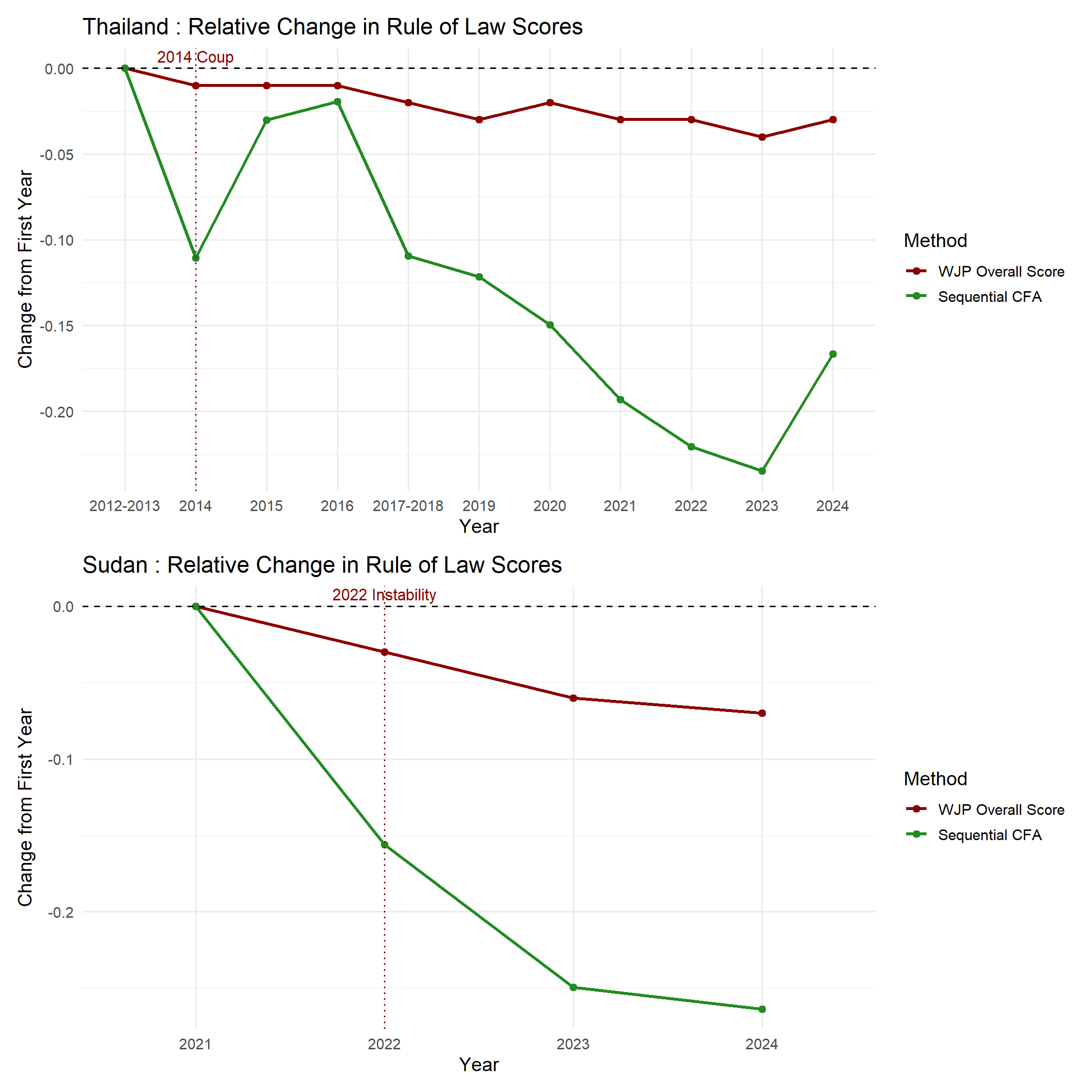}
  \caption{Comparison Between Sequential CFA and WJP ROL Scores for Thailand and Sudan}
  \label{fig:mostcomplex_pcor_boxplot}
\end{figure}

\end{document}